\documentclass[letterpaper]{article} 
\usepackage{aaai2026}  
\usepackage{times}  
\usepackage{helvet}  
\usepackage{courier}  
\usepackage[hyphens]{url}  
\usepackage{graphicx} 
\usepackage{natbib}  
\usepackage{caption} 
\usepackage{algorithm}
\usepackage{algorithmic}
\usepackage{amsmath, amsfonts}
\usepackage{booktabs}
\usepackage{subcaption}
\usepackage{xcolor}
\usepackage{tcolorbox}
\definecolor{xblue}{HTML}{1DA1F2}

\usepackage{newfloat}
\usepackage{listings}
\DeclareCaptionStyle{ruled}{labelfont=normalfont,labelsep=colon,strut=off} 
\floatstyle{ruled}
\newfloat{listing}{tb}{lst}{}
\floatname{listing}{Listing}
\title{Augmenting Intra-Modal Understanding in MLLMs for Robust Multimodal Keyphrase Generation}
\author{
    Jiajun Cao\textsuperscript{\rm 1,5}, Qinggang Zhang\textsuperscript{\rm 3}\thanks{Corresponding author.}, Yunbo Tang\textsuperscript{\rm 2}, Zhishang Xiang\textsuperscript{\rm 2}, Chang Yang\textsuperscript{\rm 3}, Jinsong Su\textsuperscript{\rm 2,4,5}\footnotemark[1]
}
\affiliations {
    \textsuperscript{\rm 1}Department of Digital Media Technology, Xiamen University  
    \textsuperscript{\rm 2}School of Informatics, Xiamen University\\
    \textsuperscript{\rm 3}The Hong Kong Polytechnic University  
    \textsuperscript{\rm 4}Shanghai Artificial Intelligence Laboratory\\
    \textsuperscript{\rm 5}Key Laboratory of Digital Protection and Intelligent Processing of Intangible Cultural Heritage of Fujian and Taiwan (Xiamen University), Ministry of Culture and Tourism, China\\
    caojiajun@stu.xmu.edu.cn, zqg.zhang@hotmail.com, jssu@xmu.edu.cn
}

\begin{document}

\maketitle

\begin{abstract}
Multimodal keyphrase generation (MKP) aims to extract a concise set of keyphrases that capture the essential meaning of paired image–text inputs, enabling structured understanding, indexing, and retrieval of multimedia data across the web and social platforms. Success in this task demands effectively bridging the semantic gap between heterogeneous modalities. While multimodal large language models (MLLMs) achieve superior cross-modal understanding by leveraging massive pretraining on image-text corpora, we observe that they often struggle with modality bias and fine-grained intra-modal feature extraction. This oversight leads to a lack of robustness in real-world scenarios where multimedia data is noisy, along with incomplete or misaligned modalities. To address this problem, we propose AimKP, a novel framework that explicitly reinforces intra-modal semantic learning in MLLMs while preserving cross-modal alignment. AimKP incorporates two core innovations: (i) Progressive Modality Masking, which forces fine-grained feature extraction from corrupted inputs by progressively masking modality information during training; (ii) Gradient-based Filtering, that identifies and discards noisy samples, preventing them from corrupting the model’s core cross-modal learning. Extensive experiments validate AimKP’s effectiveness in multimodal keyphrase generation and its robustness across different scenarios.

\end{abstract}

\begin{links}
    \link{Code}{https://github.com/XMUDeepLIT/AimKP}
\end{links}

\section{Introduction}

\begin{figure}[t]
    \centering
    \includegraphics[width=\columnwidth]{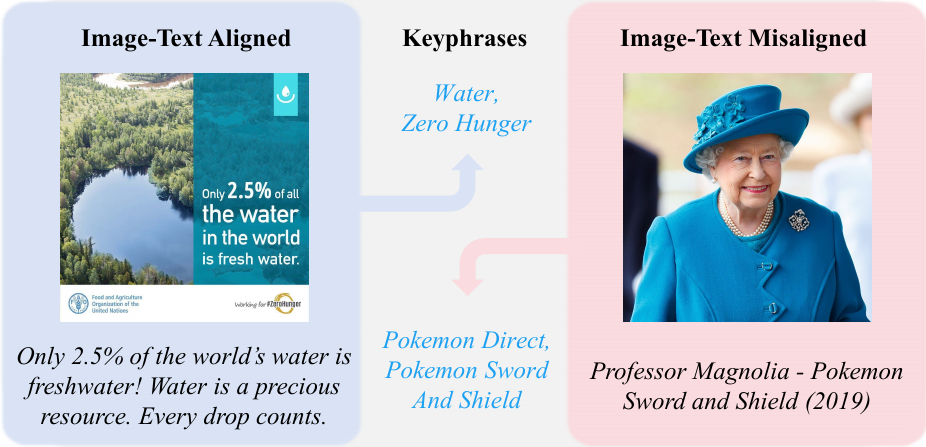}
    \caption{Examples of MKP, demonstrating cases of image-text aligned (left) and image-text misaligned (right) pairs.}
    \label{fig:example}
\end{figure}

With the explosive growth of multimedia content across the web and social platforms, there is an increasing demand for advanced techniques to understand and organize multimodal data. Multimodal keyphrase generation (MKP) addresses this critical need by generating concise, semantically rich keyphrases that encapsulate the essential meaning of multimodal inputs, enabling structured understanding, efficient indexing, and cross-modal retrieval.
For instance, consider the example in the left panel of Figure~\ref{fig:example}, where the text emphasizes global freshwater scarcity and the image depicts a freshwater lake with related slogans and organizational logos. An effective MKP system should generate both the explicit keyphrase \emph{Water} and the implicit thematic keyphrase \emph{Zero Hunger}. This capability enables critical applications such as opinion mining and content recommendation, where complementary multimodal features are needed to yield accurate and human-aligned keyphrases.

Compared to traditional text-based keyphrase generation~\cite{DBLP:conf/emnlp/ChenZ0YL18, DBLP:conf/acl/YuanWMTBHT20, DBLP:conf/acl/YeGL0Z20}, MKP requires the model to achieve both granular comprehension of modality-specific semantics for anchoring critical cues, and cross-modal integration for aligned semantic fusion.
Earlier MKP methods predominantly focus on cross-modal alignment via attention mechanisms~\cite{DBLP:conf/ijcai/GongZ16, DBLP:conf/aaai/Zhang00TY019}, frequently incorporating external tools such as OCR systems, object detectors~\cite{2020-cross-media-keyphrase-MultiHead}, or APIs~\cite{2023-Visual-Entity-Multi-granularity}. Although utilizing these auxiliary resources enhances multimodal semantic understanding, such methods are fundamentally limited by the base models' reasoning capabilities. As shown in the right panel of Figure~\ref{fig:example}, when presented with textual descriptions of Pokémon's fictional Professor Magnolia alongside images depicting the real-world Queen Elizabeth, these lightweight models struggle to disambiguate cross-modal entities, leading to erroneous keyphrase generation.

Recently, the advent of multimodal large language models (MLLMs)~\cite{DBLP:conf/nips/AlayracDLMBHLMM22,DBLP:conf/icml/0008LSH23,2023-LLaVA} has revolutionized multimodal understanding. By leveraging massive pretraining on image-text corpora, MLLMs exhibit remarkable capabilities in text recognition and visual grounding, and have set new benchmarks in tasks like image captioning and visual question answering~\cite{DBLP:journals/corr/abs-2303-08774,2024-LLaVA-v1.5,2024-qwen2.5technicalreport}.
Despite recent advances, directly deploying MLLMs for MKP is challenging due to their divergent objectives: MKP requires a fine-grained understanding of modality-specific semantics for keyphrase generation, whereas MLLMs prioritize cross-modal alignment, inherently sacrificing granular semantics.

\begin{figure}[t]
\centering
\includegraphics[width=0.9\columnwidth]{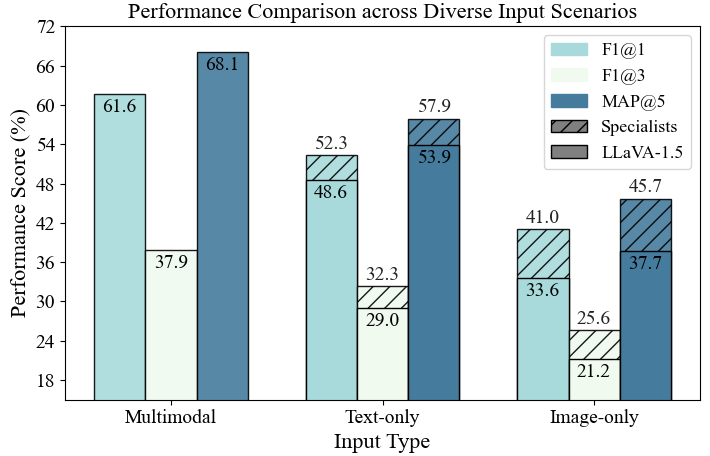}
\caption{Performance comparison of MLLMs fine-tuned on multimodal vs. unimodal contexts across three input settings: full multimodal input, text-only input, and image-only input, with metrics: F1@1, F1@3, MAP@5.}
\label{fig:comparison}
\end{figure}

This gap is quite evident in practice. 
The preliminary study on LLaVA-1.5 (Figure~\ref{fig:comparison}) shows that this representative MLLM achieves competitive performance on multimodal data, but suffers from severe degradation when processing single-modality context. More critically, it underperforms specialized single-modality models by 4\%-8\%,  with the largest discrepancy (8\%) occurring in image-only scenarios.
These observations reveal two critical points:
(i) MLLMs struggle with intra-modal understanding.
Existing MLLMs are trained on tightly aligned multimodal data, their cross-attention mechanism encourages the model to prioritize high-level cross-modal associations over fine-grained, modality-specific details. This inadvertently suppresses modality-specific reasoning capabilities to anchor keyphrases in specific visual or textual cues, which are essential for keyphrase generation.
(ii) MLLMs always suffer from modality bias. Most MLLMs exhibit a strong preference for a specific modality~\cite{DBLP:conf/iclr/ParcalabescuF25equally,DBLP:journals/corr/abs-2505-20977evaluatebias, DBLP:conf/cvpr/ZhangTSZRLYML25Debias, DBLP:journals/corr/abs-2505-18657mllmsdeeplyaffected}. For example, LLaVA exhibits a strong textual bias due to its predominantly language-based pretraining: when processing complex multimodal inputs, it tends to increase text weighting and ignore subtle visual cues.
Such imbalance violates the core requirement of MKP for adaptive modality fusion. 

These limitations become more obvious in practical scenarios, as real-world multimedia data typically exhibits noise along with incomplete or misaligned modalities.
To address this problem, we introduce AimKP, a unified training framework that adapts MLLMs for MKP through two innovations: (i) \emph{Progressive Modality Masking} that forces fine-grained feature extraction from corrupted inputs by progressively masking of modality information, and (ii) \emph{Gradient-Based Filtering} dynamically prunes masked samples based on their gradients, preventing conflicting signals from harmful corruptions. To the best of our knowledge, we are the first to propose a framework that systematically adapts MLLMs to MKP task. Our contributions are summarized as follows:

\begin{itemize}
    \item We identify key limitations of MLLMs in MKP and, based on these findings, propose AimKP, a novel framework to adapt MLLMs for the task.

    \item AimKP first introduces Progressive Modality Masking, a scheme that systematically masks modality information to force fine-grained feature extraction.
   
    \item To stabilize training, AimKP also incorporates Gradient-Based Filtering, which measures the similarity of gradients to prune uninformative or harmful masked samples.

    \item Extensive experiments on the benchmark dataset demonstrate that AimKP substantially improves MLLMs' intra-modal understanding and achieves a new state-of-the-art in overall MKP performance.
\end{itemize}

\section{Preliminary Study}
Before going into the details of AimKP, we first conduct a preliminary study to explore the potential and challenges of applying MLLMs in MKP. This study serves two core purposes: (i) to verify whether MLLMs, with their strong cross-modal alignment capabilities, can serve as a viable foundation for MKP; and (ii) to identify critical limitations in their current performance that demand targeted improvements, laying the groundwork for the design of our framework. 

\subsection{MKP with MLLMs}

To delve into MLLMs for MKP, we leverage a representative MLLM (LLaVA-1.5), which consists of a CLIP vision encoder~\cite{openclip}, a lightweight visual adapter, and a Vicuna~\cite{2023-vicuna} language model backbone.
As illustrated in Figure~\ref{fig:framework}(a), the image $X_\text{V}$ is divided into $24 \times 24$ non-overlapping patches, which are then flattened into a 1D sequence. These patches are encoded into visual embeddings via the vision encoder and adapter, with each patch’s embedding functioning as a token in the language model. Concurrently, the text $X_\text{T}$ is appended with a task prompt, tokenized, and embedded using the model's text encoder. We then concatenate the visual embeddings and the textual embeddings into a unified multimodal input following the instruction tuning paradigm. The model is trained to autoregressively generate the full keyphrase sequence $Y = \{y_1,...,y_{|Y|}\}$ conditioned on the inputs, maximizing the likelihood of ground-truth keyphrases.

For the fine-tuning setup, the vision encoder is frozen. We train the visual adapter and language model jointly, with Low-Rank Adaptation (LoRA) applied to the language model. The training loss is standard cross-entropy loss:
\begin{equation}
\mathcal{L} = -\!\underset{(X_{\text{V}},X_{\text{T}},Y)\sim\mathcal{D}}{\mathbb{E}}\Bigl[\sum_{t=1}^{|Y|}\text{log} \ p_\theta(y_{\text{t}} | X_\text{V}, X_\text{T},  y_{<t})\Bigr] .
\end{equation}

\subsection{Identifying the Intra-Modal Deficit}

We focus on evaluating MLLMs' performance across diverse scenarios, particularly focusing on how they handle modality-specific semantics.
Empirical results in Figure~\ref{fig:comparison} show that LLaVA achieves strong performance on multimodal inputs, outperforming existing MKP methods with an F1@1 of 61.6\% and MAP@5 of 68.1\%. This confirms MLLMs' capacity for cross-modal understanding. However, when either the image or the text is missing, the model's performance drops significantly. Further comparing with single-modality specialists (i.e., LLaVA fine-tuned and performing inference solely on text or images for keyphrase generation), LLaVA lags behind the text specialist by 3.7\% in F1@1 and 4.0\% in MAP@5 on text-only inputs, and 7.4\% in F1@1 and 8.0\% in MAP@5 behind the image specialist on image-only inputs, leaving its \textbf{intra-modal performance far from its theoretical ceiling}.

We attribute these results to modality bias and a critical behavioral shortcut: on well-aligned training data, the model becomes \textbf{over-reliant on cross-modal associations} because this strategy is often the easiest path to minimize loss. When its understanding of one modality is insufficient, it compensates with cues from the other rather than developing robust intra-modal understanding. Compounding its \textbf{inherent textual preference}, this combination weakens the model's ability to grasp visual details, widening the gap in image-only scenarios. This strategy, however, becomes fragile in real-world scenarios where one modality may be uninformative, noisy, or even misleading. In these cases, underdeveloped intra-modal reasoning provides no reliable fallback, leading to severe performance degradation when fusing these problematic cross-modal signals.

\begin{figure}[t]
\centering
\includegraphics[width=\columnwidth]{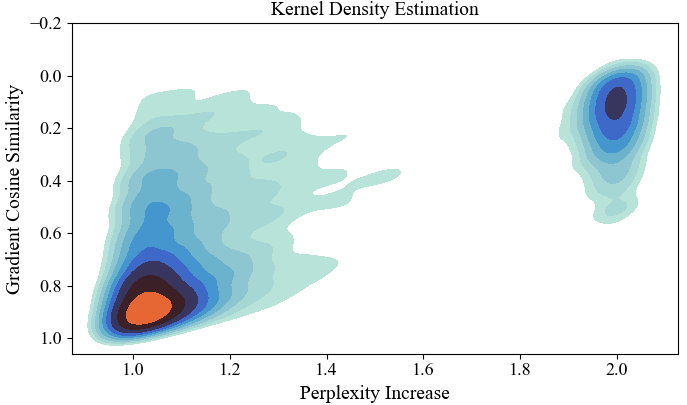}
\caption{Kernel density plot of cosine similarity of gradients vs. perplexity increase (the ratio of masked-sample keyphrase perplexity to original-sample perplexity).}
\label{fig:grad_density}
\end{figure}

\subsection{Motivation for Gradient-Based Filtering Strategy}
To address this problem, a natural intuition is to mask one modality during training, forcing the model to reason more from the unmasked modality. However, naive modality masking is risky: when core intra-modal cues are masked, the samples become uninformative noise and can undermine training. Hence, we require a mechanism that flags when a masked sample would steer the model toward divergent directions. Drawing on gradient balancing from multi-task learning~\cite{wei2024mmpareto}, we compute the cosine similarity between the gradient of the original loss and that of each masked variant. Figure~\ref{fig:grad_density} shows a clear negative relationship between gradient similarity and the increase in perplexity caused by masking. Samples whose masking barely raises perplexity (preserving key information for keyphrase generation) tend to have high gradient similarity. In contrast, those with large increases in perplexity exhibit low gradient alignment. This empirical observation supports our hypothesis that \textbf{gradient similarity can serve as a reliable flag to identify uninformative masked samples}, enabling us to filter them out and stabilize training.

\begin{figure*}[t]
    \centering
    \includegraphics[width=\textwidth]{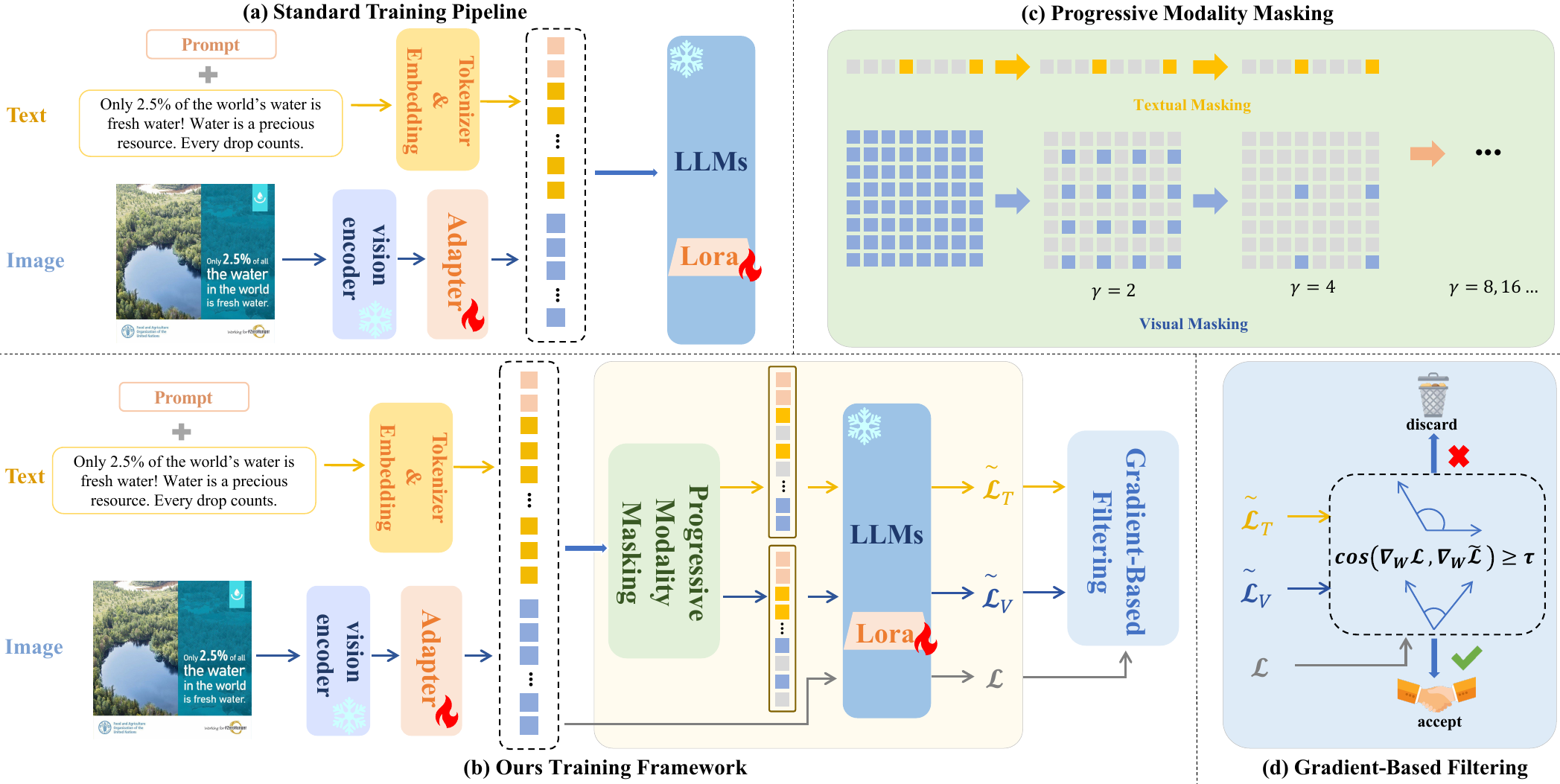}
    \caption{The Framework of AimKP. (a) Standard multimodal fine-tuning. (b) Our intra-modal enhancement framework, which (c) progressively masks modality information at increasing rates to force the model to reason deeply within one modality, and (d) dynamically prunes masked samples based on their gradients, preventing conflicting signals from harmful corruptions.}
    \label{fig:framework}
\end{figure*}

\section{The Framework of AimKP}
In this section, we propose a unified framework to enhance intra-modal learning without sacrificing cross-modal alignment. As illustrated in Figure~\ref{fig:framework}(b), AimKP comprises: (i) \emph{Progressive Modality Masking}, which forces the model to reason deeply within one modality by progressive masking information of the other modality; (ii) \emph{Gradient-Based Filtering}, which filters out uninformative masked samples, avoiding conflict with the core learning objective.

\subsection{Progressive Modality Masking}
The underdeveloped intra-modal reasoning in MLLMs arises from their modality bias and over-reliance on cross-modal associations as a training shortcut. 
To address the intra-modal reasoning deficit, we progressively mask information from one modality, compelling the model to extract rich semantics from the corrupted input. These masked samples are further filtered to retain only informative ones, as described in the next section.

For each training pair $(X_{\text{V}}, X_{\text{T}})$, we apply structured, gradually increasing masks to both image and text inputs by setting the corresponding regions of the attention mask to zero for masked areas. Both modalities undergo masking at the embedding level: text masking retains tokens at fixed intervals along the sequence, while image masking preserves tokens based on their pre-flattened spatial positions (i.e., height and width in the original grid), ensuring alignment with the 2D structure of images. Specifically,
given a stride parameter $\gamma$, we define binary masks over 2D visual features and 1D textual tokens:
\begin{equation}
\begin{aligned}
M_{\text{2D}}(i,j) &=
\begin{cases}
1, & (i \bmod \gamma = 0)\,\wedge\,(j \bmod \gamma = 0),\\
0, & \text{otherwise},
\end{cases}
\\
M_{\text{1D}}(t) &=
\begin{cases}
1, & t \bmod \gamma = 0,\\
0, & \text{otherwise}.
\end{cases}
\end{aligned}
\end{equation}
In practice, we retain the last token within each stride. For a given stride $\gamma=k$, the retention ratio is $1/k^2$ for image tokens and $1/k$ for text tokens. These masks are applied to produce masked inputs:
\[
\tilde{X}_{\text{V}} = M_{\text{2D}} \odot X_{\text{V}},
\quad
\tilde{X}_{\text{T}} = M_{\text{1D}} \odot X_{\text{T}},
\]
where ``$\odot$'' denotes token-wise masking. $\gamma$ is initialized to 2 for both modalities and doubled each epoch, systematically increasing masking intensity. This As $\gamma$ grows, the amount of retained information decreases in a controllable manner. Starting with mild masking (e.g., $\gamma=2$, retaining 50\% text tokens or 25\% image tokens) to help the model fundamentally adapt to partially missing inputs; then as masking intensifies (e.g., $\gamma=4$, retaining 25\% text tokens or 6.25\% image tokens), the model is forced to mine gradually deeper intra-modal understanding. This aligns with curriculum learning principles, where the model adapts to progressively more challenging intra-modal reasoning tasks. We further introduce refinements based on sample informativeness to $\gamma$ detailed in the following section.

\subsection{Gradient–Based Filtering}
While progressive masking enhances the model's intra-modal learning and escalates task difficulty, overly aggressive masking can introduce noise or remove essential signals about the task, destabilizing training. As noted in our preliminary analysis, gradient similarity serves as an effective proxy for the informativeness of a given masked sample. To filter out uninformative variants, we compute the cosine similarity between the gradient of the original loss $\mathcal{L}$ and that of its masked counterpart $\mathcal{\tilde{L}}$:
\begin{equation}
s = \cos\bigl(\nabla_{W}\mathcal{L},\,\nabla_{W}\mathcal{\tilde{L}}\bigr).
\end{equation}
High similarity score suggests the masking is effective and produces informative variants, whereas low similarity suggests the masked sample introduces conflicting gradients that can harm the optimization of the primary objective $\mathcal{L}$.
We apply a threshold $\tau$ to decide each variant’s fate:
\begin{itemize}
  \item If $s \ge \tau$, we include the loss of the masked inputs as an auxiliary loss in the training objective. This high similarity can also be interpreted as an opportunity to challenge the model further, so we double the corresponding stride for the next epoch, further intensifying masking.
  \item If $s < \tau$, we set the auxiliary loss weight to zero, excluding the over-masked variant and halve the stride for the next epoch (minimum 2) to reduce the masking intensity.
\end{itemize}
This dynamic adjustment filters out samples with conflicting gradients, ensuring that only masked variants aligned with the primary objective contribute to training. The adaptive masking intensity regulation further allows us to strike a fine balance between exploring the model's intra-modal potential and avoiding unproductive training. Ultimately, this combination enhances the models' intra-modal abilities without compromising its core learning on the complete inputs.

\subsection{Training Objective}

Let $\mathcal{L}$ be the loss for generating target keyphrases $Y$ from the original, unmasked image-text inputs $(X_{\text{V}}, X_{\text{T}})$; we define two auxiliary losses to reinforce intra-modal learning. The first, ${\mathcal{\tilde{L}}}_{\text{V}}$, requires the model to generate the target $Y$ from the masked visual features $\tilde{X}_{\text{V}}$ and the complete text features ${X}_{\text{T}}$. The second, ${\mathcal{\tilde{L}}}_{\text{T}}$, is analogous, computed using $({X}_{\text{V}},\tilde{X}_{\text{T}})$. The full objective combines the original loss with the two auxiliary losses, each controlled by a dynamic 0-1 switch $\lambda_{\text{V}}$ and $\lambda_{\text{T}}$:
\begin{equation}
\mathcal{L}_{\text{total}}
= \mathcal{L}
+ \lambda_{V}\,\mathcal{\tilde{L}}_{\text{V}}
+ \lambda_{T}\,\mathcal{\tilde{L}}_{\text{T}}.
\end{equation}
We apply separate thresholds $\tau_{V}$ and $\tau_{T}$ for the image-masked and text-masked variants, the switches $\lambda_{\text{V}}$ and $\lambda_{\text{T}}$ are indicator functions determined by the gradient similarity score $s$ and the threshold $\tau$:
\[
\lambda_{V} = \mathbf{1}\{s_{V}\ge\tau_{V}\},
\quad
\lambda_{T} = \mathbf{1}\{s_{T}\ge\tau_{T}\},
\]
where $\mathbf{1}\{\cdot\}$ is the indicator function (1 if true, 0 otherwise).

\section{Experiment}
\subsection{Experiment Setup}
\label{section:setting}

\paragraph{Datasets}
Following previous studies~\cite{2020-cross-media-keyphrase-MultiHead,2023-Visual-Entity-Multi-granularity}, we carry out experiments on the CMKP dataset collected by~\cite{2020-cross-media-keyphrase-MultiHead}. This dataset consists of 53,701 English tweets collected from Twitter, each containing a unique text-image pair with user-generated hashtags as keyphrases, and split into 8:1:1 train-val-test sets.

\paragraph{Evaluation Metrics}
Following~\cite{2020-cross-media-keyphrase-MultiHead, 2023-Visual-Entity-Multi-granularity}, we adopt identical evaluation metrics:
(i) \textbf{F1@K}: Macro-F1 score for the top-K keyphrase predictions,
(ii) \textbf{MAP@K}: mean average precision on top-K predictions.
For scenarios where the model generates $n < K$ keyphrases, we pad the remaining $(K-n)$ positions with empty labels $\emptyset$ when computing F1 scores, and dynamically set $K'=\min(n, K)$ during the calculation of MAP scores. Notably, in sequential generation settings, predictions are ordered by their decoding sequence rather than confidence scores. Specifically, the order in which keyphrases are generated directly serves as their rank.

\paragraph{Baselines}
To validate the effectiveness of AimKP, we conduct a comprehensive comparison against a range of strong baselines. These baselines are divided into two main groups. First, we benchmark against existing models specifically designed for or adapted to MKP, namely \textbf{CO-ATT}~\cite{DBLP:conf/ijcai/ZhangWHHG17}, \textbf{FLAVA}~\cite{DBLP:conf/cvpr/SinghHGCGRK22}, \textbf{M\(^3\)H-ATT}~\cite{2020-cross-media-keyphrase-MultiHead}, \textbf{MM-MKP}~\cite{2023-Visual-Entity-Multi-granularity}, and text-only models adapted to MKP by leveraging image-associated text \textbf{BART-large}~\cite{DBLP:conf/emnlp/WolfDSCDMCRLFDS20} and \textbf{CopyBART}~\cite{DBLP:journals/tjs/YuGZ24}. Second, and serving as our main comparison models, we include powerful MLLMs under standard fine-tuning, specifically \textbf{LLaVA}-1.5-7B~\cite{2024-LLaVA-v1.5} and \textbf{Qwen2-VL}-7B~\cite{wang2024Qwen2vlenhancingvisionlanguagemodels}.

\paragraph{Implementation Details}
Our primary experimental setup is centered on the \textbf{LLaVA}-1.5-7B model, which we fine-tune using LoRA and optimized using Adam~\cite{DBLP:journals/corr/KingmaB14}. 
In the final loss formulation, the original loss and the two auxiliary losses are equally weighted after filtering. We have also experimented with alternative weighting strategies, but observed no consistent improvement across settings. For the gradient-based filtering, we set modality-specific thresholds $\tau_V = 0.4$ and $\tau_T = 0.1$. To establish a foundational capability, we first train the model for one epoch on normal data, and then apply progressive modality masking and the gradient-based filtering.
The training is conducted on four NVIDIA A6000 GPUs with a learning rate of 2e-4, and a total batch size of 64. We perform validation at the end of each epoch and select the model checkpoint that yields the best composite score on the validation set for final testing. All experiments are performed with three random seeds, and we report the averaged results
During inference, we decode outputs using beam search with sampling with a beam size of 5 and a temperature of 0.5, repeating the decoding process three times and taking the average of the results. Prior to evaluation, both the generated predictions and ground-truth keyphrases are stemmed with the Porter Stemmer~\cite{DBLP:journals/program/Porter06} and subsequently deduplicated. We also conduct additional experiments implementing AimKP on \textbf{Qwen2-VL}-7B to validate the effectiveness of our method across different architectures. 

\subsection{Main Results}

\begin{table}[t]
\centering
\begin{tabular}{llll}
  \toprule
  Models & F1@1 & F1@3 & MAP@5 \\
  \midrule
  \multicolumn{4}{c}{\textit{Image-only}}\\
  \midrule
  LLaVA (image specialist)  & 41.02 &25.64 &45.66\\
  LLaVA                       & 33.57 & 21.17 & 37.68 \\
  LLaVA-AimKP               & 37.41  & 23.77  & 41.94  \\
  \midrule
  \multicolumn{4}{c}{\textit{Text-only}} \\
  \midrule

  LLaVA (text specialist)     & 52.33 &32.34 &57.93\\
  LLaVA                      & 48.56 & 28.99 & 53.92 \\
  CopyBART                        &49.67 &33.89 &53.95\\
  LLaVA-AimKP               & 50.04 & 30.63  & 55.45  \\
  \midrule
  \multicolumn{4}{c}{\textit{Multimodal}} \\
  \midrule
  CO-ATT           & 42.12    & 31.55    & 48.39    \\
  FLAVA       	&46.05	&31.23	&49.30 \\
  M\(^3\)H-ATT & 47.06 & 33.11 & 52.07 \\
  MM-MKP           & 48.19 & 33.86 & 53.28 \\
  BART-large  &50.47 &34.69 &55.11\\
  CopyBART                   & 51.42 & 36.54 & 57.35\\
  LLaVA                     & 61.58 & 37.90 & 68.07 \\
  LLaVA-AimKP          & \textbf{63.16} & \textbf{39.00} & \textbf{69.96} \\
  Qwen2-VL                      & 63.08 & 38.43 & 69.89 \\
  Qwen2-VL-AimKP    & \textbf{64.18}  & \textbf{38.73}  & \textbf{71.00}  \\
  \bottomrule
\end{tabular}
\caption{Performance on image-only, text-only, and multimodal inputs on the CMKP dataset. “Specialist” denotes models fine-tuned exclusively on a single modality. AimKP refers to models trained under our framework.}
\label{tab:performance comparison on CMKP}
\vspace{-2mm}
\end{table}

\paragraph{General MKP}
Table~\ref{tab:performance comparison on CMKP} presents comparative results on the CMKP test dataset, yielding two key observations:
First, standard fine-tuned MLLMs (e.g., LLaVA-1.5, Qwen2-VL) outperform small models by a significant margin. For instance, LLaVA-1.5 achieves 61.58\% F1@1 (+10.16\%) and 68.07\% MAP@5 (+10.72\%) over the strongest CopyBART baseline. This underscores MLLMs’ powerful inherent visual understanding and cross-modal integration capabilities, while highlighting the critical role of their pretrained cross-modal knowledge in MKP.
Second, even on already high-performing MLLMs, AimKP yields consistent gains: LLaVA-1.5-AimKP improves F1@1 by 1.58\% (63.16\% vs. 61.58\%) and MAP@5 by 1.89\% (69.96\% vs. 68.07\%), while Qwen2-VL-AimKP gains 1.10\% in F1@1 and 1.11\% in MAP@5. These results validate that progressive masking combined with gradient-based filters strengthens MLLMs’ general keyphrase generation performance.
\paragraph{Intra-Modal Understanding}
Despite the gains on general MKP, AimKP also mitigates the significant gap remaining in single-modality performance. On text-only inputs, AimKP boosts MAP@5 from 53.92\% to 55.45\%, shrinking the deficit relative to the text specialist (57.93\%) from 4.01 to 2.48 points. Similarly, on image-only inputs, LLaVA-1.5’s MAP@5 score rises from 37.68\% to 41.94\% with AimKP, reducing the gap to the image specialist (45.66\%) from 7.98 to 3.72 points. These improvements, particularly the larger gains in image-only scenarios, demonstrate that AimKP enables the model to continuously deepen its intra-modal reasoning while alleviating its inherent textual bias.

Moreover, AimKP yields smaller gains on F1@3: only 1.1\% on LLaVA-1.5 and 0.3\% on Qwen2-VL. Even the text specialist (32.34\%) fails to outperform CopyBART (33.89\%) on text-only inputs. We hypothesize that stronger models tend to generate only high-certainty keyphrases, and incomplete information further worsens this, as MLLMs' advanced capabilities make them more sensitive to such incompleteness. Since the F1@3 metric penalizes outputs with fewer than three keyphrases by padding with $\emptyset$ tokens, these cautious behaviors result in lower scores.

\subsection{Ablation Study}
To validate the effectiveness and analyze the contributions of each component within AimKP, we conduct a series of ablation studies. The results are summarized in Table~\ref{tab:ablation}.

\begin{table}
\centering
\begin{tabular}{llll}
  \toprule
   Models & F1@1 & F1@3 & MAP@5 \\
  \midrule
  AimKP         & \textbf{63.16} & \textbf{39.00} & \textbf{69.96}  \\
  \midrule
  w/o masking on text           & 62.41  & 38.37  & 69.12  \\
  w/o masking on image          &62.74 &38.66 &69.43 \\
  w/o gradient-based filtering      &62.79 &38.75 &69.47 \\
  fixed masking ($\gamma=2$)   & 63.11 & 38.91 & 69.93 \\
  fixed masking ($\gamma=4$)     & 63.15 & 38.61 & 69.93
\\\bottomrule
\end{tabular}
\caption{Ablation study of AimKP's key components, w/o denotes removing this component.}
\label{tab:ablation}
\vspace{-2mm}
\end{table}

\paragraph{Effectiveness of Bimodal Masking}
Having established that our progressive masking strategy effectively improves performance, we now seek to verify the importance of applying this augmentation to both modalities. To this end, we conduct experiments where we disable masking on either the text or image inputs. The results in lines 3-4 reveal a significant performance degradation in both scenarios. Specifically, removing intra-modal augmentation of the image modality (w/o masking on text) or the text modality (w/o masking on image) causes a drop of 0.84 and 0.53 points in MAP@5, respectively. This difference aligns closely with our earlier observation of MLLMs’ inherent textual preference, which leaves image understanding relatively underdeveloped. The consistent performance drops when masking is removed from either modality confirm that our intra-modal augmentation strategy is effective for both modalities, validating its ability to strengthen modality-specific semantics.

\paragraph{Effectiveness of Gradient-Based Filtering}
The gradient-based filtering is designed to prevent noisy or counterproductive updates from harming the training process. To validate its effectiveness, we disable it by setting the threshold $\tau=-1$, which accepts all masked variants regardless of their gradient similarity. Line 5 shows that this leads to a notable performance drop across all metrics (e.g., F1@1 decreases from 63.16\% to 62.79\%). This result indicates that naively applying aggressive masking can introduce samples with corrupted semantics that generate conflicting gradients. Our filter is crucial for identifying and discarding these harmful updates, thereby safeguarding the primary learning objective and ensuring stable performance gains.

\paragraph{Progressive Masking vs. Fixed Masking}
We also compare our progressive masking strategy against fixed masking ($\gamma=2$ and $\gamma=4$) throughout training. As shown in lines 6-7, our progressive approach consistently outperforms both fixed strategies across all metrics and exhibits greater stability. An easy ratio may not provide a sufficient training signal, while a consistently hard one can introduce excessive noise. Our progressive scheme dynamically adapts the difficulty, demonstrating the benefit of a curriculum-like approach to intra-modal learning.

\begin{figure*}[t]
    \centering
    \includegraphics[width=\textwidth]{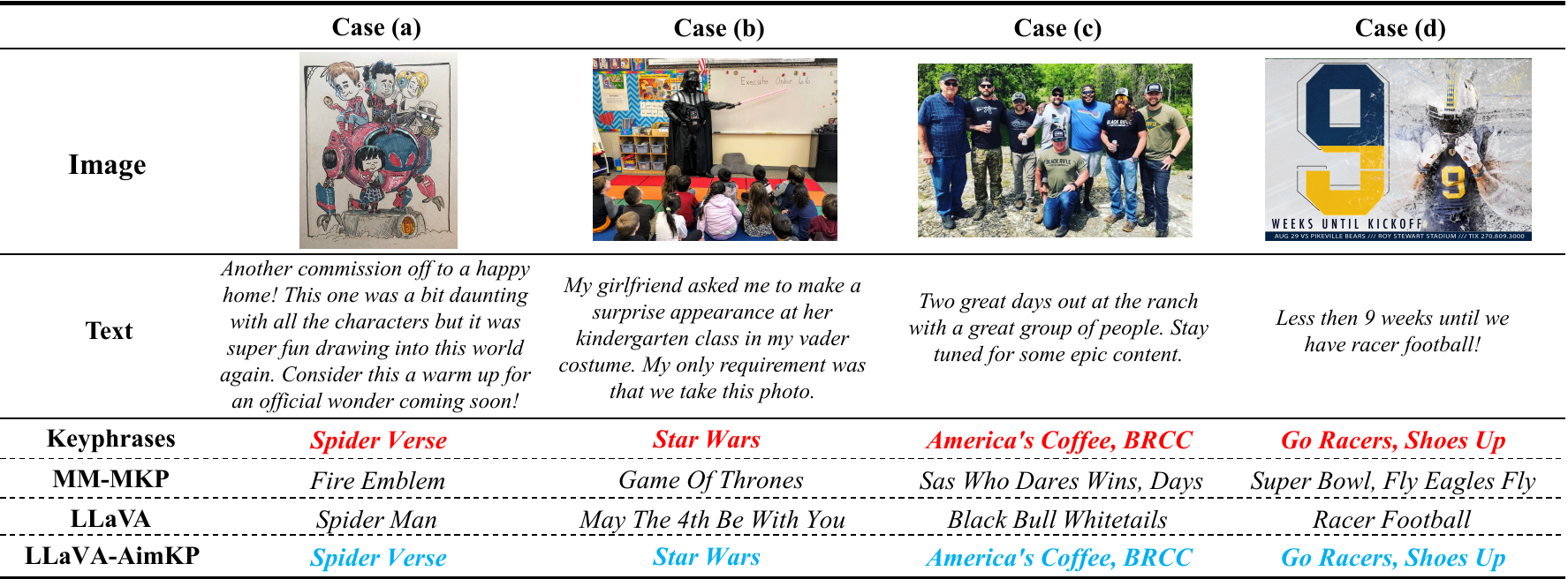}
    \caption{Case study comparing keyphrase outputs of MM-MKP, LLaVA, and LLaVA-AimKP on four examples.}
    \label{fig:cases}
\end{figure*}

\subsection{Case Study}
To further validate the effectiveness of AimKP, we compare its performance with the baseline models MM-MKP and LLaVA using selected cases in Figure~\ref{fig:cases}. 

As shown in Case (a), MM-MKP misidentifies the animated character ensemble as \emph{Fire Emblem}, while LLaVA and LLaVA-AimKP recognize \emph{Spider Man}-related themes, showcasing MLLMs’ edge in world knowledge utilization. LLaVA-AimKP further recognizes the specific character ensemble and contextual clues unique to \textbf{\emph{Spider Verse}} (i.e., Spider Man Universe). Similarly, in Case (b), both LLaVA and LLaVA-AimKP link ``vader costume'' to \textbf{\emph{Star Wars}} lore, unlike small models lacking such contextual awareness. However, LLaVA fails to deeply grasp intra-modal information and generates \emph{May The 4th Be With You} (referring to a specific fan holiday), which is absent from the inputs.

Case (c) highlights LLaVA-AimKP's fine-grained feature extraction. The inputs depict a community gathering for brand promotion, where the key cues lie in shirts with the ``Black Rifle Coffee Company'' (BRCC) logo in the image. LLaVA-AimKP, benefiting from enhanced intra-modal feature extraction, accurately identifies these brand-related visual cues and generates \textbf{\emph{America’s Coffee, BRCC}}. In contrast, LLaVA outputs \emph{Black Bull Whitetails}, a phrase clearly unrelated to any content. Furthermore, case (d) indicates that AimKP empowers the model to incorporate fine-grained features into more semantically complete predictions. The input counts down to the Racer football season opener, indicating 9 weeks until the game. While LLaVA superficially identifies \emph{Racer Football} with the text, LLaVA-AimKP captures the underlying context and emotion using information from both modalities, generating the right keyphrases \textbf{\emph{Go Racers, Shoes Up}} (the team's spirit and motto).

\section{Related Work}
\paragraph{Multimodal Large Language Models}
MLLMs~\cite{DBLP:conf/nips/AlayracDLMBHLMM22,DBLP:conf/icml/0008LSH23,DBLP:journals/corr/abs-2303-08774,2023-LLaVA} have demonstrated strong capabilities in visual question answering, image captioning, and cross-modal reasoning. Building on these foundations, recent works~\cite{team2023gemini,DBLP:conf/iclr/Zhu0SLE24,DBLP:conf/acl/LanNMLZS25,DBLP:conf/iclr/LiZZZLLML25} extend MLLMs to more complex visual tasks, improving fine-grained understanding~\cite{zhang2024knowgpt, zhang2025faithfulrag} and reasoning over intricate scenes~\cite{xiang2025use}.

\paragraph{Multimodal Keyphrase Generation}
Keyphrase generation (KPG) has been a significant area of focus within natural language processing~\cite{zhuang2025linearrag}. Current neural KPG models can be broadly categorized into three paradigms~\cite{DBLP:journals/ipm/XieSSWWYLXS23}:
(i) \textsc{one2one}~\cite{DBLP:conf/emnlp/ChenZ0YL18, DBLP:conf/acl/MengZHHBC17, DBLP:conf/aaai/ChenGZKL19}, which converts a training sample containing multiple keyphrases into several training instances. Each instance pairs the source input with a single keyphrase.  
(ii)~\textsc{one2seq}~\cite{DBLP:conf/acl/YuanWMTBHT20, DBLP:conf/acl/ChenCLK20, DBLP:conf/naacl/KulkarniMAB22} treats KPG as a sequence-to-sequence task and concatenates all ground-truth keyphrases into a single target sequence according to a predefined order.
(iii) \textsc{one2set}~\cite{DBLP:conf/acl/YeGL0Z20, DBLP:conf/emnlp/XieWYLXWZS22, DBLP:conf/emnlp/ShaoZPMYSS24}, which models KPG as a set generation problem, generating keyphrases as an unordered set in parallel.
While these works focus primarily on text-based KPG, a growing body of work is exploring the multimodal domain. Common approaches use co-attention networks to integrate text and visual information~\cite{DBLP:conf/ijcai/GongZ16, DBLP:conf/aaai/Zhang00TY019}.~\citet{2020-cross-media-keyphrase-MultiHead} incorporates explicit optical characters and implicit image attributes from external tools, developing a MKP encoder-decoder model with multi-head attention mechanism.~\citet{2023-Visual-Entity-Multi-granularity} further enhances textual inputs with visual entities from external APIs and mitigates image noise through multi-granularity filtering.

\paragraph{Intra-Modal Augmentation}
Intra-modal augmentation addresses modality imbalance by strengthening models' comprehension on each modality. A common approach frames this as multi-task learning, applying unimodal losses to each encoder. For instance, Self-MM~\cite{yu2021learning} generates dynamic unimodal labels as auxiliary supervision; UMT~\cite{du2021improving} employs a teacher–student framework, combining fusion loss and distillation loss to align each encoder with a unimodal teacher. Taking a different approach,
EAU~\cite{gao2024embracing} learns unimodal representations by explicitly modeling data uncertainty during contrastive learning. Others focus on balancing the optimization process directly. OGM-GE~\cite{peng2022balanced} and MMPareto~\cite{wei2024mmpareto} dynamically reweight gradients from the primary and auxiliary losses to correct imbalances.

To the best of our knowledge, we are the first to adapt MLLMs for MKP while mitigating modality bias and over-reliance on cross-modal shortcuts. Unlike existing methods which use static unimodal losses or focus on task-level gradient balancing, we introduce progressive modality masking to dynamically raise intra-modal difficulty, and sample-level gradient-based filtering to retain only informative masked samples. This ensures MLLMs build robust intra-modal understanding while maintaining cross-modal strengths.

\section{Conclusion}
In this paper, we introduce AimKP, a novel framework that addresses inherent modality bias and underdeveloped intra-modal understanding of MLLMs in MKP. AimKP leverages progressive modality masking to compel fine-grained feature extraction within each modality, and employs gradient-based filtering to remove uninformative masked samples, thereby stabilizing the training process. Extensive experiments and analyses demonstrate that AimKP not only strengthens MLLMs' intra-modal capabilities but also achieves a new state-of-the-art in overall performance.

Future work will focus on adapting AimKP to other multimodal tasks where modality imbalance similarly degrades performance and investigating transition learning~\cite{DBLP:journals/pami/ZhouLMZXWZS23} to better leverage diverse modalities.
\section{Acknowledgements}
The project was supported by National Key R\&D Program of China (No.2022ZD0160501), Natural Science Foundation of Fujian Province of China (No.2024J011001), and the Public Technology Service Platform Project of Xiamen (No.3502Z20231043). We also thank the reviewers for their insightful comments.
\bibliography{aaai2026}
\appendix

\twocolumn[ 
  \centering
  \LARGE\bfseries Appendix\\ 
  \vspace{0.5cm} 
]
\section{A. Experimental Setup} 
\subsection{A.1 Dataset}
CMKP dataset includes 53,701 English tweets, each of which comprises a distinct text-image pair, with user-annotated hashtags serving as keyphrases. Table~\ref{tab:dataset} characterizes the dataset across splits, text complexity, keyphrase density, and vocabulary diversity.

\begin{table}[ht]
\centering
\begin{tabular}{lccccc}
\toprule
\textbf{Split} & \textbf{Size} & \textbf{Text Len} & \textbf{$\vert\text{KP}\vert\text{/s}$} & \textbf{$\vert\text{KP}\vert$} & \textbf{KP Len} \\
\midrule
Train & 42,959 & 27.26 & 1.33 & 4,261 & 1.85 \\
Valid & 5,370 & 26.81 & 1.34 & 2,544 & 1.85 \\
Test & 5,372 & 27.05 & 1.32 & 2,534 & 1.86 \\
\bottomrule
\end{tabular}
\caption{CMKP Dataset Statistics.
\textbf{Text Len}: Average token count in input text.
\textbf{$\vert\text{KP}\vert\text{/s}$}: Average number of keyphrases per sample.
\textbf{$\vert\text{KP}\vert$}: Total distinct keyphrases in the split.
\textbf{KP Len}: Average token length of keyphrases.}
\label{tab:dataset}
\end{table}

\subsection{A.2 Implementation Details}
We use the instruction format shown in Figure~\ref{fig:instruction} and adopt the training hyperparameters listed in Table \ref{tab:hyper}. For comparative fairness, we use the same hyperparameters for Qwen2-VL as for other models, except that it requires more steps to converge. Thus, we set the number of epochs to 8 for standard fine-tuning and 6 for Qwen2-VL-AimKP. Since Qwen2-VL employs dynamic image resolution while LLaVA fixes the number of image tokens to 576, we constrain the number of image tokens in Qwen2-VL to approximately 576. During inference on Qwen2-VL, multi-beam search yields suboptimal results, so we adopt a sampling strategy with beam size 1. To compute per-sample gradients for gradient-based filtering, we set the per-GPU batch size to 1 with 16 gradient accumulation steps. AimKP adds ~6 hours to the ~9-hour baseline training time while inference times are identical.
\begin{figure}[ht]
\centering
\begin{tcolorbox}[
    colback=gray!5!white, 
    colframe=gray!50!black, 
    coltitle=white,
    fonttitle=\bfseries\normalsize,
    fontupper=\small,
    title=Instruction Template
]
\emph{[System-Message]}\\
\textbf{USER}: \emph{[Image][Text]}\textbackslash nWhat phrases should be used to tag the media?\\
\textbf{ASSISTANT}: \emph{[Keyphrase A], [Keyphrase B], ...}
, however, yields higher sample efficiency: the baseline required 5× data to match fit, whereas AimKP achieves better results with less compute (1×normal + 3×under AimKP).

\end{tcolorbox}
\vspace{-4pt}
\caption{Prompt template.}
\label{fig:instruction}
\end{figure}

\begin{table}[ht]
\centering
\begin{tabular}{l|cc}
\toprule
Hyperparameter    & LLaVA & LLaVA-AimKP  \\
LoRA              & \multicolumn{2}{c}{$r=128, \alpha=256$} \\
Epoch             & 6& 4\\
Batch size        & \multicolumn{2}{c}{64} \\
LoRA lr           & \multicolumn{2}{c}{2e-4} \\
Adapter lr        & \multicolumn{2}{c}{2e-5}   \\
lr schedule       & \multicolumn{2}{c}{cosine decay}      \\
lr warmup ratio   & \multicolumn{2}{c}{0.03}              \\
Weight decay      & \multicolumn{2}{c}{0}                 \\
Optimizer         & \multicolumn{2}{c}{AdamW}             \\
DeepSpeed stage   & 2       & -           \\ \bottomrule
\end{tabular}
\caption{Hyperparameters used in training.}
\label{tab:hyper}
\end{table}

\subsection{A.3 Baseline Models}
\textbf{M\(^3\)H-ATT} uses a multimodal encoder to process text and visual content, while enhancing inputs with OCR text and image attributes (nouns/adjectives) via external tools. Features are integrated through a multi-head attention module, and then sent to the prediction module. The module combines keyphrase classification and generation, with a pointer network to copy words from source inputs, and the final output dynamically balances generated and copied results.

\noindent\textbf{MM-MKP} builds on M³H-ATT with architectural refinements. It incorporates visual entities into the text stream via external APIs and enhances image processing with multi-granularity denoising, leveraging global text-image similarity and regional attention to focus on key visual areas. Training follows a two-stage paradigm: pre-training with matching and classification losses, and then fine-tuning with combined classification and generation loss.

\noindent\textbf{CopyBART}, originally a text-based keyphrase generation model built on BART, adapts to multimodal scenarios by extending text inputs with image attributes and OCR text. It uses a ``One2MultiSeq'' dual-order training paradigm (training on both original and reversed keyphrase sequences) for data augmentation. The model employs a copy mechanism  in the decoder to copy words from textual inputs and balance generation and copying.

\section{B. Additional Ablation Studies}
\subsection{B.1 Alternative strategies}
We conducted additional comparisons to further validate the design choices of AimKP, with results in Table~\ref{tab:appendix_ablation}.  

\begin{table}[ht]
\centering
\begin{tabular}{llll}
  \toprule
   Models & F1@1 & F1@3 & MAP@5 \\
  \midrule
  AimKP         & \textbf{63.16} & \textbf{39.00} & \textbf{69.96}  \\
  \midrule
  W/o warm up         & 62.99 & 38.90 & 69.74 \\
  Random masking           & 62.83  & 38.77  & 69.59  \\
  Linear increase	         &62.96	&38.67	&69.76 \\
  Feature compression      & 62.83 & 38.74 & 69.45 \\
  \bottomrule
\end{tabular}
\caption{Additional ablation results comparing training strategy, masking pattern, and information reduction methods. \emph{w/o warm up} means applying progressive modality masking and gradient-based filtering from start, \emph{linear increase} indicates strides increasing at linear rate.}
\label{tab:appendix_ablation}
\vspace{-2mm}
\end{table}

In our setup, we introduce progressive modality masking and gradient-based filtering after a warm up training on normal data for one epoch to establish models' basic instruction following and task understanding in the initial phase. Compared to applying these mechanisms from the start (line 2), this initialization strategy not only accelerates training but also yields better performance. Introducing masking-enhanced tasks too early may overwhelm the model with excessive difficulty, hindering the development of foundational capabilities. 

We also further compare our structured masking with random masking and linear increase (ie. $\gamma=1,2,3 ...$). While both of them fail to guarantee strictly stronger masking because masks may not be nested, structured masking ensures controlled information reduction as $\gamma$ increases and outperforms them (line 3, 4).  

Finally, we tested an alternative information reduction method: feature compression via pooling instead of masking. This approach compresses features to reduce information but fails to retain original positional relationships. As shown in line 5, compression performs worse than masking, indicating that preserving positional information is critical.

\subsection{B.2 Thresholds Sensitivity}
Thresholds $\tau_T$ and $\tau_V$ were chosen via validation-set sweeps, the method is not sensitive within a reasonable range.  
\begin{table}[ht]
\centering
\begin{tabular}{llll}
  \toprule
   $\tau_T$, $\tau_V$ & F1@1 & F1@3 & MAP@5 \\
  \midrule
    0.1, 0.4       &\textbf{63.3}	&\textbf{39.0}	&\textbf{70.3} \\
  \midrule
    0.0, 0.1	&63.0	&38.6	&69.9\\
    0.05, 0.15	&63.0	&39.1	&70.0\\
    0.1, 0.3	&63.2	&38.8	&70.1\\
    0.1, 0.5	&62.8	&38.7	&69.8\\
  \bottomrule
\end{tabular}
\caption{Ablations on validation set show that AimKP is not sensitive to thresholds $\tau_T$ and $\tau_V$}
\label{tab:appendix_ablation_T}
\vspace{-2mm}
\end{table}

\subsection{B.3 Cost\&Data Augmentation}
AimKP adds ~6 hours to the ~9-hour baseline training time (4 x A6000), but only during offline training; inference times are identical. Importantly, AimKP yields higher sample efficiency: the baseline required roughly 5× data to match fit, whereas AimKP can achieves better results~\ref{tab:appendix_ablation_C} with less effective total compute (1× normal training + 3× under AimKP).

\begin{table}[ht]
\centering
\begin{tabular}{llll}
  \toprule
   Models & F1@1 & F1@3 & MAP@5 \\
  \midrule
    LLaVA &61.7 &37.4 &68.2\\
    AimKP &62.0 &37.8 &68.8\\
  \bottomrule
\end{tabular}
\caption{Ablations on data efficiency, AimKP achieves better results even with less total data.}
\label{tab:appendix_ablation_C}
\vspace{-2mm}
\end{table}

\section{C. Training Analysis}  
We visualize the evolution of gradient similarity metrics during training, derived from the progressive modality masking and gradient-based filtering process.

\begin{figure}[ht]
    \centering
    \includegraphics[width=\columnwidth]{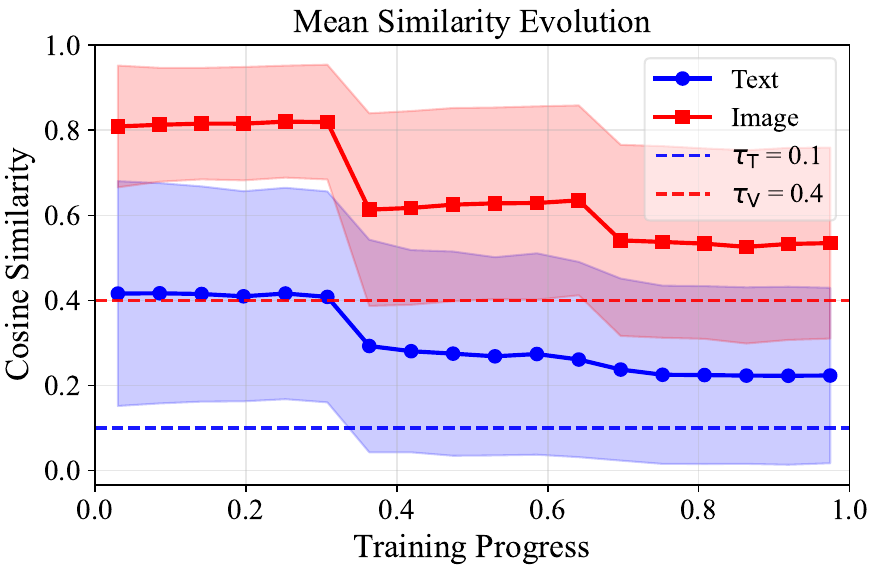}
    \caption{Mean cosine similarity between masked and normal samples, with light-colored shaded areas representing standard deviation. \emph{Text} denotes text-masked samples, and \emph{Image} denotes image-masked samples.}
    \label{fig:mean_similarity}
\end{figure}

Figure~\ref{fig:mean_similarity} displays the mean cosine similarity between masked and normal samples.  We can observe that image-masked samples consistently exhibit higher gradient similarity compared to text-masked samples throughout training. We attribute this phenomenon to the inherent redundancy of the image modality, which means even with increased masking intensity (larger $\gamma$), sufficient critical information remains preserved. This redundancy allows the model to maintain relatively consistent gradient updates between masked and normal image samples, resulting in higher similarity scores. In contrast, text modality relies on compact, sequential token dependencies, where masking key tokens can more easily disrupt semantic integrity, leading to lower gradient similarity for text-masked samples.

\begin{figure}[ht]
    \centering
    \includegraphics[width=\columnwidth]{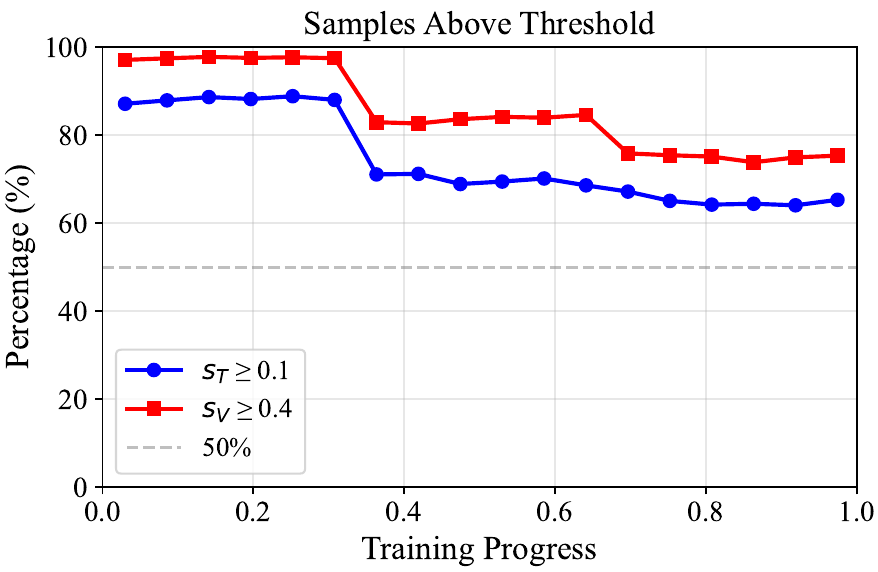}
    \caption{Percentage of samples above thresholds}
    \label{fig:above_threshold}
\end{figure}

As training progresses and masking intensity increases (stride $\gamma$ grows), the overall gradient similarity shows a downward trend. Correspondingly, the percentage of valid samples (above thresholds) decreases: for text-masked samples, it drops from 87\% to 62\%, and for image-masked samples, from 96\% to 77\%. This trend indicates that stronger masking introduces more challenging samples, leading to fewer valid instances as $\gamma$ increases. Notably, the slowing rate of decline in the later stages of training suggests that our halving method is effective: instead of rigidly increasing masking intensity for all samples, it dynamically adjusts $\gamma$ to find the optimal stride for samples that cannot tolerate further masking, thereby stabilizing the training process.

\section{D. Supplementary Examples}
\subsection{D.1 Additional Case Study}  
\begin{figure*}[ht]
    \centering
    \includegraphics[width=\linewidth]{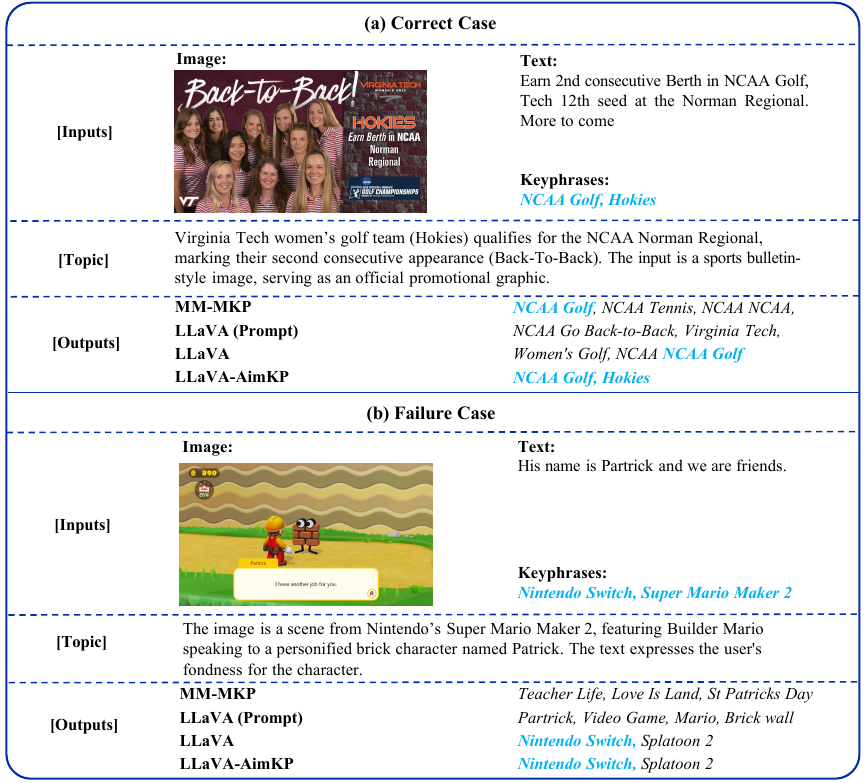}
    \caption{Examples of additional MKP cases. 
    (a) A successful case with target keyphrases \textbf{\emph{NCAA Golf, Hokies}}. The untuned LLaVA generates related but overly general phrases such as \emph{Virginia Tech, Women's Golf, NCAA}, while the standard fine-tuned LLaVA extracts only \textbf{\emph{NCAA Golf}}. In contrast, LLaVA-AimKP precisely outputs all target keyphrases. 
    (b) A failure case where the correct keyphrase is \textbf{\emph{Nintendo Switch, Super Mario Maker 2}}. The untuned LLaVA produces only loosely related concepts like \emph{Video Game, Mario, Patrick}, while both the fine-tuned LLaVA and LLaVA-AimKP incorrectly generate \textbf{\emph{Nintendo Switch}, }\emph{Splatoon 2}, highlighting a common challenge in disambiguating specific named entities.}
    \label{fig:supplementary_cases}
\end{figure*}

We present additional case studies (both correct and failure cases) in Figure~\ref{fig:supplementary_cases}, including outputs from the base \textbf{LLaVA}-1.5-7b model using an MKP-specific prompt. We observe that the untuned LLaVA is capable of generating contextually relevant phrases, but often fails to produce the precise keyphrases required. This limitation is particularly pronounced when the keyphrases involve abbreviations (e.g., ``LFC'' for Liverpool Football Club) or domain-specific terminology (e.g., ``TX Lege'' for the Texas Legislature), both of which frequently appear in the CMKP dataset, leading to poor performance under standard metrics.

In the failure case Figure~\ref{fig:supplementary_cases}(b), the image depicts a scene from \textbf{\emph{Super Mario Maker 2}} featuring the well-known Super Mario and a character named Patrick, and the text expresses fondness for Patrick. Both LLaVA and LLaVA-AimKP capture the keyphrase \textbf{\emph{Nintendo Switch}} but incorrectly generate \emph{Splatoon 2} (another Nintendo game) instead of the target \textbf{\emph{Super Mario Maker 2}}. This highlights lingering limitations of current models in disambiguating domain-specific entities and aligning multimodal cues accurately.

\subsection{D.2 Training Examples}
\begin{figure*}
    \centering
    \includegraphics[width=\linewidth]{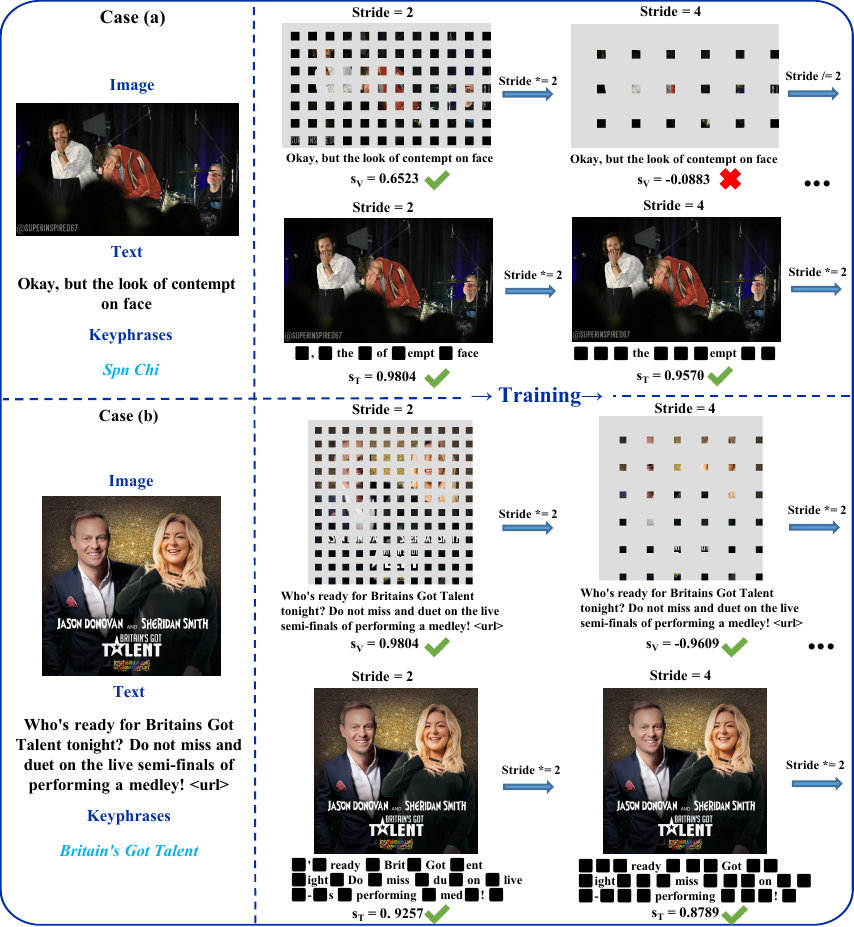}
    \caption{Training examples of Progressive Modality Masking and Gradient-Based Filtering. Image/text modality masking guided by gradient similarity scores $(s_V, s_T)$. Samples are accepted for training and masking is intensified $(stride *= 2)$ when scores exceed thresholds $(\tau_T=0.1, \tau_V=0.4)$; otherwise, samples are pruned and the stride is reduced $(stride /= 2)$. In case (a), task-relevant information (e.g., the fan convention scene of Supernatural in Chicago) is primarily contained in the image. Thus, masking the text has little impact but excessive masking of the image directly disrupts the core information, leading to a sharp drop in similarity. In case (b), as the image and text share redundant information (both promoting Britain's Got Talent, a talent show),  masking either modality leaves sufficient information for learning, making both masking strategies viable.}
    \label{fig:training_cases}
\end{figure*}

Figure~\ref{fig:training_cases} illustrates concrete instances of Progressive Modality Masking and Gradient-Based Filtering during training, showcasing how masking intensity (stride) adjusts dynamically across samples.

\end{document}